\documentclass[journal]{IEEEtran}
\usepackage{cite}
\usepackage[rightcaption]{sidecap}

\ifCLASSINFOpdf
   
   \usepackage[pdftex]{graphicx}
   \usepackage{multirow}
   \usepackage{tabularew}
   \usepackage[flushleft]{threeparttable}
   
   \graphicspath{{./pdf/}{./jpeg/}{./eps/}}
  \DeclareGraphicsExtensions{ ,.jpeg,.png}
\else
   \usepackage[dvips]{graphicx}
   \graphicspath{{./eps/}}
   \DeclareGraphicsExtensions{.eps}
\fi
\usepackage{amsmath}
\usepackage{array}

\ifCLASSOPTIONcompsoc
 \usepackage[caption=false,font=normalsize,labelfont=sf,textfont=sf]{subfig}
\else
  \usepackage[caption=false,font=footnotesize]{subfig}
\fi

\usepackage{stfloats}
\usepackage{url}
  \usepackage{color}

\usepackage{url}

\begin{document}

\title{Xcel-RAM: Accelerating Binary Neural Networks in High-Throughput SRAM Compute Arrays}

\author{Amogh~Agrawal*, Akhilesh~Jaiswal*, Deboleena~Roy, Bing~Han, Gopalakrishnan~Srinivasan, Aayush~Ankit, Kaushik~Roy,~\IEEEmembership{Fellow,~IEEE}
\thanks{The authors are with the School of Electrical and Computer Engineering,
Purdue University, West Lafayette, IN-47907, USA}
\thanks{(* These authors contributed equally)}
}

\maketitle
\pagenumbering{gobble}

\begin{abstract}

Deep neural networks are biologically-inspired class of algorithms that have recently demonstrated state-of-the-art accuracy in large scale classification and recognition tasks. Hardware acceleration of deep networks is of paramount importance to ensure their ubiquitous presence in future computing platforms. Indeed, a major landmark that enables efficient hardware accelerators for deep networks is the recent advances from the machine learning community that have demonstrated the viability of aggressively scaled deep binary networks. In this paper, we demonstrate how deep binary networks can be accelerated in modified von-Neumann machines by enabling binary convolutions within the SRAM array. In general, binary convolutions consist of bit-wise XNOR followed by a population-count (\textit{popcount}). We present two proposals $-$ one based on charge sharing approach to perform vector XNORs and approximate \textit{popcount} and another based on bit-wsie XNORs followed by a digital bit-tree adder for accurate \textit{popcount}. We highlight the various trade-offs in terms of circuit complexity, speed-up and classification accuracy for both the approaches. Few key techniques presented as a part of the manuscript is use of low-precision, low overhead ADC, to achieve a fairly accurate \textit{popcount} for the charge-sharing scheme and proposal for \textit{sectioning} of the SRAM array by adding switches onto the read-bitlines, thereby achieving improved parallelism. Our results on a benchmark image classification dataset CIFAR-10 on a binarized neural network architecture show energy improvements of 6.1$\times$ and 2.3$\times$ for the two proposals, compared to conventional SRAM banks. In terms of latency, improvements of 15.8$\times$ and 8.1$\times$ were achieved for the two respective proposals.

\end{abstract}

\begin{IEEEkeywords}
In-memory computing, SRAM, binary convolution, binary neural networks, deep-CNNs.
\end{IEEEkeywords}

%
\IEEEpeerreviewmaketitle

\section{Introduction}

Deep convolutional neural networks (CNNs) have been established as the state-of-the-art for recognition and classification tasks \cite{bengio2009learning,jones2014learning}, often surpassing human capabilities \cite{silver2016mastering,He_2016,gonature}. Most popular networks that won the ImageNet \cite{imagenet} challenge, such as AlexNet \cite{krizhevsky2012imagenet}, GoogLeNet \cite{googlenet}, ResNet \cite{He_2016}, \textit{etc.,} are based on deep CNNs. However, hardware running these networks consume large amounts of energy, in fact, orders of magnitude more than the human brain \cite{schneider2018ultralow}. This immense energy-gap stems from the underlying architecture of the current state-of-the-art hardware implementations, that are variants of the von-Neumann machines \cite{vnbottleneck}. They contain physically separate computation and memory blocks, connected via a system bus. Although this architecture has worked wonders for general-purpose computing tasks, when it comes to deep CNNs and data intensive applications in general, frequent data transfers between the memory and the computation unit becomes a bottleneck, given the limited bandwidth of the bus. Moreover, since each transaction is expensive, a large power penalty is incurred per memory access.

Recent developments in the neural network community have identified these problems and have come up with simpler \textit{memory-friendly} algorithms. Binary neural networks \cite{binarybengio,abhrobnn,srinivasan2016magnetic} and XNOR-nets \cite{rastegari2016xnor} have been recently developed and shown large potential. The idea is to reduce the precision of input activations and the network weights to single-bit. This immensely simplifies the computations to Boolean bit-wise operations, with only minimal degradation in the state-of-the-art accuracies. Since convolution is the most power-hungry operation in neural networks, it is reduced to a bit-wise XNOR followed by a population count (\textit{popcount}) of the XNORed output. This opens pathways for adopting new simplified binary \textit{in-memory} computing paradigms for accelerating neural networks. As shown in \cite{agrawal2017x,6tddc,8tdpe}, bit-wise Boolean operations including XORs or XNORs as well as non-Boolean vector-matrix dot-products can easily be incorporated within standard SRAM arrays. Such SRAM based \textit{in-memory} computations open up new possibilities of augmenting the existing memory arrays with compute capabilities. Thereby, one can imagine a modified von-Neumann machine, which can cater well to general purpose computing tasks as well as act as on-demand compute accelerator. 

To that effect, we propose novel techniques to compute in-memory binary convolutions, as an added functionality to the standard 10-transistor (T) SRAM bitcells. In the first approach (Proposal-A), we use charge-sharing between the parasitic capacitances present inherently in the SRAM array to perform the XNOR and \textit{popcount} operations involved in the binary convolution. Although this approach is digital, with binary weights and binary inputs stored in the memory array, the \textit{popcount} is generated as an analog voltage on the source-lines. In order to sense this analog voltage, we propose a low-overhead and low-precision ADC (owing to area and energy constraints in the memory array).
Another key highlight of this approach is that we employ a \textit{sectioned-SRAM} by dividing memory sub-banks into smaller sections. With n-sections in a particular sub-bank we can accomplish n-binary convolutions in parallel. This is important because obtaining the \textit{popcount} output for large kernels is non-trivial. For large networks, the kernel sizes in deeper layers are typically too large to be stored in a single row of a given memory sub-array. As such, \textit{popcount} for larger binary networks inevitably requires a scheme to estimate the partial \textit{popcount} from each row, which can then be summed up from different sub-arrays to get the final \textit{popcount}. 
However, the low-overhead and low-precision ADC induces approximations in the \textit{popcount} output, which results in overall system accuracy degradation. Thus, we propose another approach (Proposal-B), where we alter the peripheral circuitry of the SRAM array and enable two word-lines simultaneously. This approach, although not as energy/throughput efficient as Proposal-A, generates accurate XNOR and \textit{popcount} operations, thereby, not affecting the overall system accuracy. The proposed circuit techniques in Proposal-A and Proposal-B allows us to process multiple kernels at once, thereby improving the overall system throughput and making the proposal suitable for a range of deep binary networks.


There have been several earlier works to develop hardware platforms that can accelerate CNN algorithms. Hardware architectures that use highly sub-banked memory units feeding an array of multiply-accumulate processing elements have been presented in many works including \cite{chen2014dadiannao,chen2017eyeriss,agrawal_spare}. A key drawback of such distributed processing array based customized design is the fact that it makes the underlying computing hardware application specific and in many cases specific to neural network accelerators. Further, emerging technologies like memristive crossbars have been employed in many proposals as convolution accelerators geared towards neural networks in general \cite{resparc,isaac}. The very use of memristors as convolution engine renders such platforms unsuitable for general purpose computing due to various challenges faced by memristive state-of-the-art technologies. These include, the limited endurance of memristive devices, the multi-cycle write-verify programming scheme \cite{memristors_prob1} and the drift in programmed resistance state with aging \cite{chen2011variability}. 
More recently, \cite{biswas2018conv} demonstrated an analog approach to binary convolution using charge-sharing. However, the work presented in \cite{biswas2018conv} was limited to smaller networks. This is because with larger and more complex networks, the inaccuracies in interfacing conventional low precision DACs/ADCs unacceptably degrades the network accuracy. 



The main highlights of the present work are as follows:

\begin{enumerate}

\item We present two novel techniques to compute binary convolutions. Proposal-A uses charge-sharing between the parasitic capacitances inherently present within the standard 10T-SRAM array, to accomplish a fairly accurate \textit{popcount} operation. Proposal-B alters the SRAM peripheral circuitry to perform accurate in-memory XNOR and \textit{popcount} operations. 


\item Further, we propose \textit{sectioned-SRAM} to increase parallelism within the SRAM arrays, thereby improving the computation throughput and energy-efficiency of the binary convolution operation. 

\end{enumerate}

\begin{figure}[t]
\centering
\includegraphics[width=0.35\textwidth]{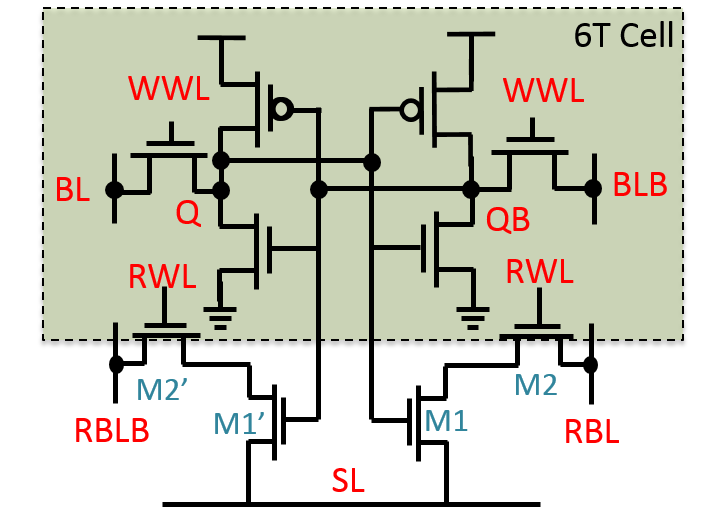}
\caption{The 10 transistor SRAM cell featuring a differential decoupled read-port comprising of transistors M1-M2 and M1'-M2'. The write port is constituted by write access transistors connected to WWL.}
\centering
\label{fig:10t}
\end{figure}

\section{In-memory Binary Convolution $-$ Proposal-A}

As discussed in the introduction, a convolution operation is simplified to a bitwise XNOR, followed by a \textit{popcount} of the XNORed output in binary neural networks (BNNs). Although bitwise XNOR operation is simple to incorporate within the memory, the \textit{popcount} operation is not very straightforward. We exploit the inherent SRAM structure, utilizing the internal parasitic capacitances to perform the XNOR and \textit{popcount} of two vectors stored within the memory array. Although our approach to binary convolution is digital, we sense an analog voltage within the memory array to evaluate the \textit{popcount} output. Sensing analog voltages in general, is difficult without precise ADCs. Most common precise ADCs, such as Flash ADCs and SAR type ADCs require excessively large power and area \cite{razavi_adc}, making them unsuitable for memory applications. Thus, we propose a dual read-wordline (Dual-RWL) along with a dual-stage ADC to minimize the errors in the \textit{popcount} output. Further, we describe the sectioned-SRAM technique to improve the throughput and the energy-efficiency of the binary convolution. Since the same set of inputs need to be convolved with multiple kernels, each section in sectioned-SRAM stores a different kernel while the inputs are shared among all sections, thereby performing the operations concurrently.

\subsection{Circuit Description}

We use a standard 10T-SRAM cell as the basic memory unit. Fig. \ref{fig:10t} shows a schematic of the 10T-SRAM cell, containing the basic 6T-cell as the storage unit, along with transistors M1-M2 and M1'-M2' forming the differential read ports, respectively. Writing into the cell is functionally similar to the 6T write operation through the write-ports (WWL, BL, BLB). For reading, RBL and RBLB are pre-charged to $V_{DD}$, SL is connected to ground, and RWL is enabled. If the bit-cell stores a `1' (Q = $V_{DD}$, QB = 0V), RBL discharges to 0V and RBLB holds its charge. Similarly, if the bit-cell stores a `0' (Q = 0V, QB = $V_{DD}$), RBLB discharges to 0V and RBL holds its charge. A differential sense amplifier senses the voltage difference between RBL and RBLB to generate the output.

\begin{figure*}[t]
\centering
\includegraphics[width=\textwidth]{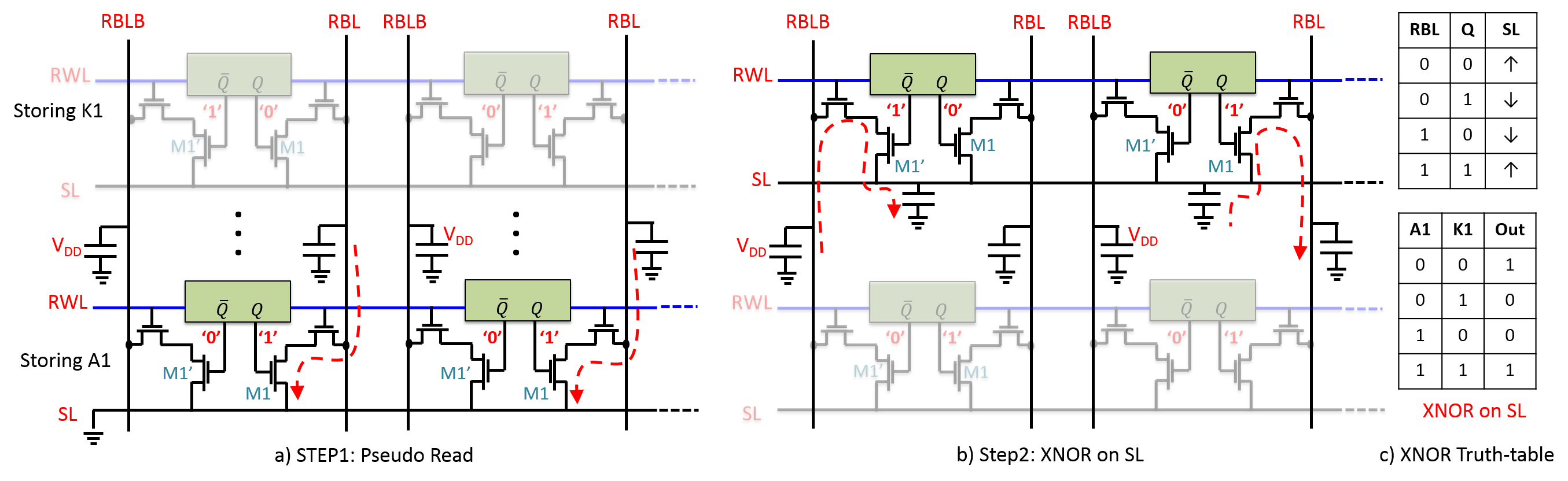}
\caption{Illustration of the binary convolution operation within the 10T-SRAM array. a) Step 1: Pseudo-read. RBLs/RBLBs are pre-charged and RWL for a row storing the input activation (A1) is enabled. Depending on the data A1, RBLs/RBLBs either discharge or stay pre-charged. The SAs are not enabled, in contrast to a usual memory read. Thus, the charge on RBLs/RBLBs represent the data A1. b) Step 2: XNOR on SL. Once the charges on RBLs/RBLBs have settled, RWL for the row storing the kernel (K1) is enabled. Charge sharing occurs between the RBLs/RBLBs and the SL, depending on the data K1. The RBLs either deposit charge on the SL, or take away charge from SL. c) The truth table for Step 2 is shown. The pull-up and pull-down of the SL follow the XNOR truth table. Moreover, since the SL is common along the row, the pull-ups and pull-downs are cumulative. Thus, the final voltage on SL represents the XNOR + \textit{popcount} of A1 and K1.}
\centering
\label{fig:binconv}
\end{figure*}

\begin{figure*}[t]
\centering
\includegraphics[width=\textwidth]{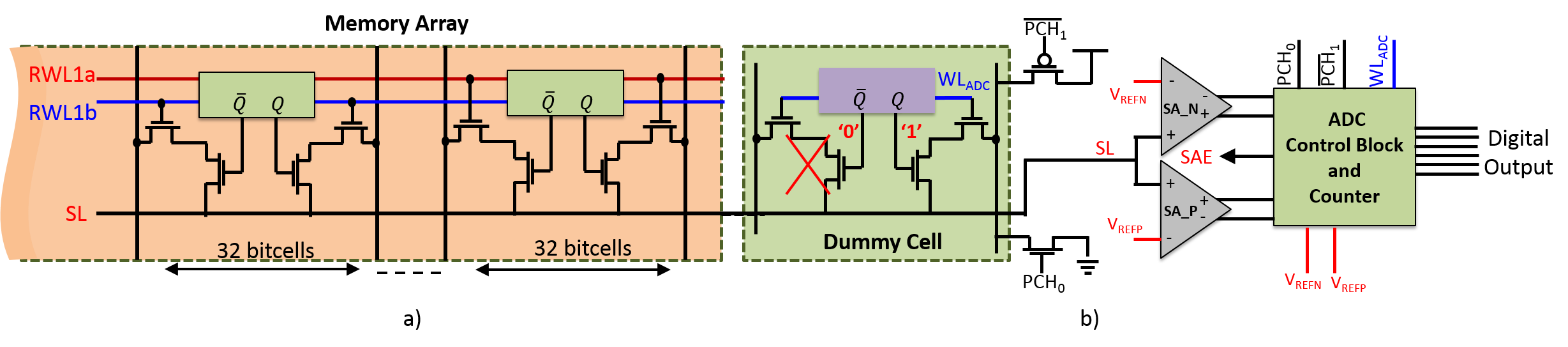}
\caption{a) Dual RWL technique. The left block shows the memory array with Dual RWL. Each row in the memory array consists of two read-wordlines RWL1 and RWL2. Half of the cells along the row are connected to RWL1, while the other half are connected to RWL2. At a time, only one of RWL1, RWL2 are enabled, to ensure that only half of the cells participate in charge sharing, at a time, thereby reducing the number of voltage states on the SL to be sensed. b) Dual-stage ADC scheme. The ADC consists of two dummy bitcells (only one shown), two SAs, counter and a control block. The ADC control block generates the reference signals $V_{REFN}$ and $V_{REFP}$, and SAE, which are fed to the two SAs. These are used in the first-stage of the ADC to determine the sub-class (first 2bits of ADC output). It also generates the signals $PCH_0$, $\overline{PCH_1}$, $WL_{ADC}$ which operate on the dummy cells during the second-stage, to either pump-in or pump-out  charge from SL, depending on the sub-class. The counter counts the number of cycles during the process to generate the final 3bits of the ADC output.}
\centering
\label{fig:adc}
\end{figure*}

We use the inherent parasitic capacitances on RBLs, RBLBs and SLs (C$_{RBL}$, C$_{RBLB}$ and C$_{SL}$, respectively) in the 10T-SRAM structure to compute the binary convolution within the memory array itself. The operation can be described in three steps as follows: 

\paragraph{\textbf{Pseudo-read}} A read operation is performed on a row storing the binary vector inputs, say A1 (refer Fig. \ref{fig:binconv}(a)). First, all RBLs/RBLBs are precharged to $V_{DD}$, as in the usual read operation. Next, when the RWL corresponding to the row storing A1 is enabled, the precharged RBLs and RBLBs discharge conditionally, depending on the data values, thereby stabilizing at $V_{DD}$ or 0V. For the example shown in the figure, the data stored is `1' in both cells corresponding to the input vector A1, thus, both RBLs discharge to 0V and RBLBs stay at $V_{DD}$. Note that the differential sense-amplifiers are not enabled in this \textit{pseudo-read} step.

\paragraph{\textbf{XNOR on SL}} After the pseudo-read operation, the RBLs/RBLBs store the information of A1 as their respective voltages. Now, the RWL of the row storing a weight kernel, say K1, is enabled (refer Fig. \ref{fig:binconv}(b)). Interestingly, this causes charge-sharing between C$_{RBL}$, C$_{RBLB}$ and C$_{SL}$ as shown in the figure by the charge current paths. In the example, the two cells corresponding to K1 store a `0' and `1' respectively. Thus, when the RWL is enabled, charge flows into the SL from M1' in the left cell, while charge flows out of SL through M1 in the right cell. This `pull-up'($\uparrow$) and `pull-down'($\downarrow$) of the SL follows the XNOR operation of the data stored in the cell (K1) and the RBL/RBLB charge (A1). With respect to the example chosen above, one can observe that the first two rows of the XNOR truth table of Fig. \ref{fig:binconv}(c) are taken care-of. If the bits corresponding to the activation (A1) was `0' and `0', \textit{i.e.,} RBLB is at 0V while RBL is at $V_{DD}$, then the charge flows out of SL through M1' in left cell, while it flows into SL through M1 in right cell. This represents the bottom two rows of the XNOR truth table. Thus, we perform a bitwise XNOR operation between vectors A1 and K1, represented by the charge stored on the line SL.

\paragraph{\textbf{Popcount}} Since the SL is shared by all the cells along the row, these `pull-ups' and `pull-downs' are cumulative. As can be seen from Fig. \ref{fig:binconv}(c), an SL `pull-up' corresponds to a `1' in the output XNORed vector, while an SL `pull-down' corresponds to a `0' in the output XNORed vector. In order to evaluate the \textit{popcount} of the output vector, we need to count the number of 1's. More 1's in the output vector implies more `pull-ups' on SL, which in turn implies a higher voltage on SL. Thus, the final SL voltage represents the \textit{popcount} of the output vector: (A1 XNOR K1).  We boost the RWL voltage such that the SL swing is from 0V to $V_{DD}$. To sense this analog voltage we use a charge-sharing based sequentially integrating ADC, adopted from \cite{biswas2018conv}. Note however, that this is an approximate, low-precision ADC. Thus, in order to achieve a fairly accurate estimate of the \textit{popcount} of the entire row, we use two techniques described in the next sub-section. 

\subsection{Dual Read-Wordline based Dual-stage ADC}

In order to evaluate the \textit{popcount} of the entire memory row at once, we should be able to distinguish N- number of distinct states in the analog SL voltage, where N is the number of columns in a memory array (we choose N=64, for a reasonably sized array). In the output XNORed vector, there can zero 1's, one 1's, two 1's, ... , up to N 1's. Correspondingly, there are N- different voltage levels on the SL, which need to be sensed by the ADCs. However, due to area and power constraints within the memory array, it is infeasible to use high-precision ADCs, such as the area-expensive SAR or power-hungry Flash ADCs. We adopt a simple charge-sharing based serially integrating type ADC for our purposes. However, instead of having to sense N- distinct analog levels, the ADC only needs to sense N/8 levels. This is enabled by using a Dual RWL memory structure, along with a dual-stage ADC. 

\begin{figure}[t]
\centering
\includegraphics[width=0.5\textwidth]{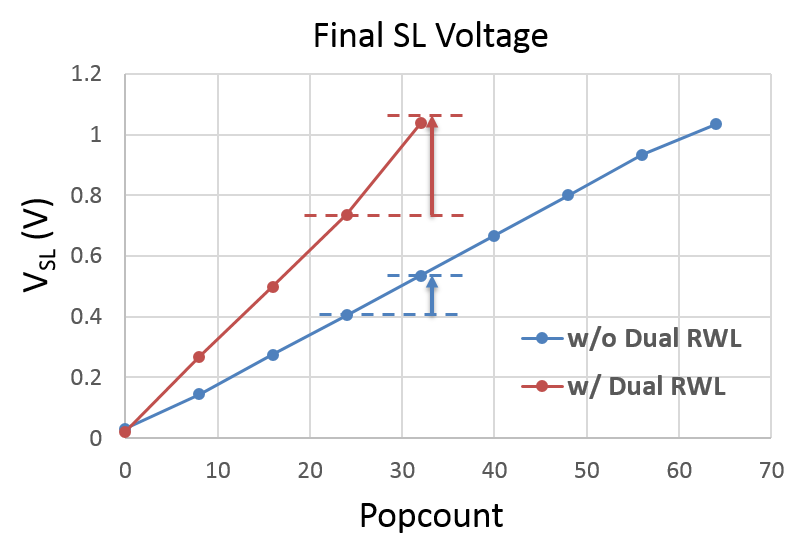}
\caption{The plot shows the final SL voltage with and without the Dual RWL approach. A larger sense margin is obtained with our Dual RWL approach, thus relaxing the constraints on the low-overhead ADC. Note that with Dual RWL technique we restrict the distinct voltage levels on SL to 32 at a time, instead of 64. However, the voltage swing on SL remains the same, thereby increasing the sense margin between the states.}
\centering
\label{fig:slresults}
\end{figure}

The Dual RWL technique is shown in Fig. \ref{fig:adc}(a). Note that we use two sets of read word-lines (RWL1a, RWL1b) for every memory row. First half of the cells along the row are connected to RWL1a, while the rest are connected to RWL1b. The step 2 of the binary convolution (XNOR on SL) described above is split in two parts. First, only RWL1a is enabled. Thus, only N/2 cells are enabled to share charge with SL, either pulling-up or pulling-down the SL voltage. The rest half of the cells are cut off from the SLs, and cannot participate in the charge sharing. Once the SL voltage has been sensed by the ADC, RWL1a is disabled, and RWL1b is enabled. Now, the other half of the cells share charge to generate a voltage on SL. Note that this does not change the swing on the SL, since the SL voltage depends on the capacitive ratio $C_{RBL}$/$C_{SL}$. Thus, the N/2 voltage levels are equally separated out from 0V to $V_{DD}$. This can be confirmed from Fig. \ref{fig:slresults}, which shows the SL voltages for N=64, with and without Dual RWL technique, as a function of the \textit{popcount}. Since the separation between the states has increased, it becomes easier to sense the levels with a low-overhead ADC.

\begin{figure}[t]
\centering
\includegraphics[width=0.5\textwidth]{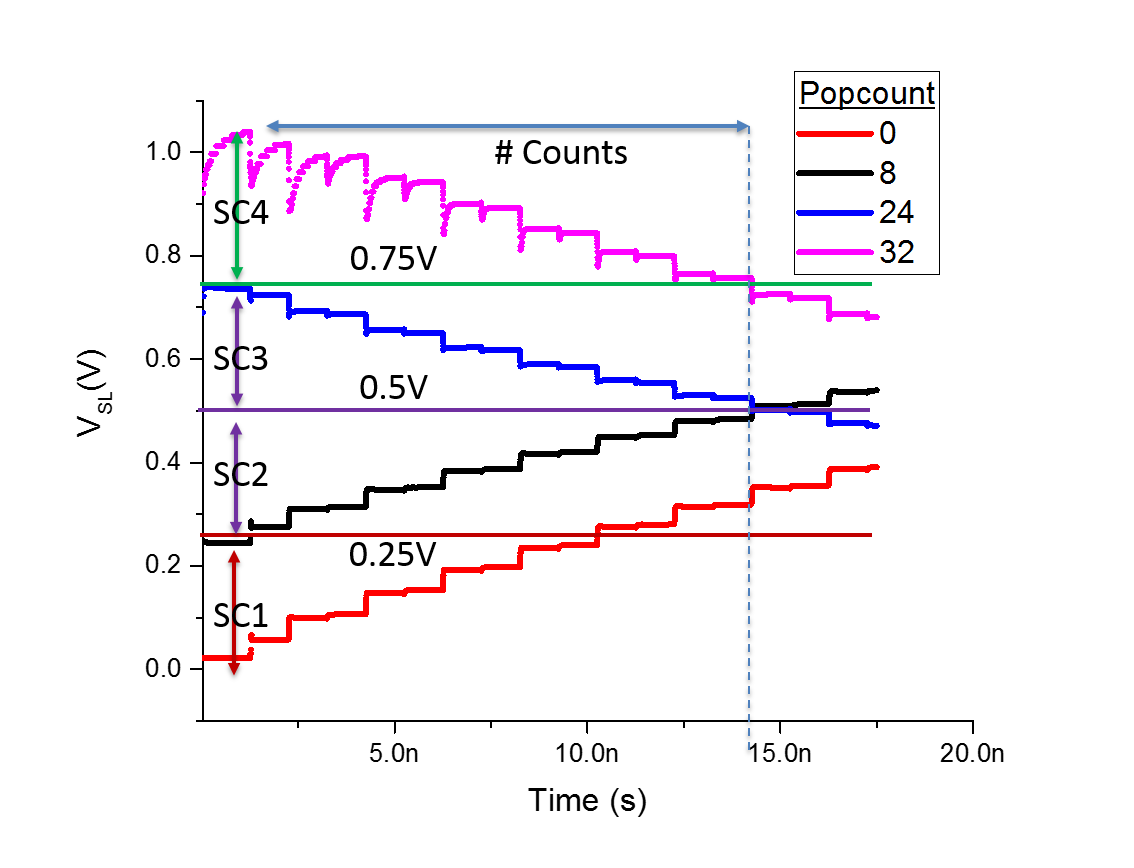}
\caption{The figure shows the timing diagrams for the dual-stage ADC scheme. The figure plots the SL voltage for various \textit{popcount} cases. In the first-stage, the sub-class SC1-4 is determined using multiple references (0.25V, 0.5V and 0.75V). In the second-stage, charge is pumped-in/out of SL successively, depending on the SC. The number of cycles it takes for SL to reach $V_{REF}$ are counted. $V_{REF}$ for SC1-4 is 0.25V,0.5V, 0.5V and 0.75V, respectively.}
\centering
\label{fig:steps}
\end{figure}

The ADC used is shown schematically in Fig. \ref{fig:adc}(b). It consists of two dummy bitcells per row (only 1 shown in figure), two SAs, a counter and an ADC logic block. We employ a dual-stage ADC to sense the analog voltage on SL. In the first-stage for ADC sensing, we use multiple voltage references ($V_{DD}$/4, $V_{DD}$/2 and 3$V_{DD}$/4), to classify the analog voltage levels into four sub-classes SC1, SC2, SC3, SC4 $-$ [0-$V_{DD}$/4], [$V_{DD}$/4-$V_{DD}$/2], [$V_{DD}$/2-3$V_{DD}$/4] and [3$V_{DD}$/4-$V_{DD}$], respectively. This is done using two voltage SAs, since the voltage swing on SL spans 0V to $V_{DD}$. On SA\_N, a V$_{REF}$ of 3$V_{DD}$/4 is applied, while for SA\_P, a V$_{REF}$ of $V_{DD}$/4 is applied. If both SA outputs are LOW, the SL voltage is classified in SC1. Similarly, when both SA outputs are HIGH, the SL voltage is classified in SC4. Otherwise, V$_{REF}$ is changed to $V_{DD}$/2, and the SA outputs are observed again. If both outputs are HIGH, the SL voltage is classified in SC3, otherwise SC2. Thus, the first-stage of the ADC generates the MSB 2bits of the ADC output.

Once the sub-classes of the analog voltage have been defined, the second-stage of the ADC is initiated. The ADC logic block generates a bunch of control signals $-$ $PCH_0$, $\overline{PCH_1}$ and WL$_{ADC}$, which operate on the dummy bitcells. For SC1 and SC2, SA\_P is enabled with a V$_{REF}$ of $V_{DD}$/4 and $V_{DD}$/2, respectively. $\overline{PCH_1}$ is pulsed alternately with WL$_{ADC}$, to pump-in a small amount of charge into SL every cycle through the dummy cells. In each cycle, when WL$_{ADC}$ is LOW and $\overline{PCH_1}$ is HIGH, the RBL of the dummy cell is precharged to $V_{DD}$. When WL$_{ADC}$ is HIGH and $\overline{PCH_1}$ is LOW, the precharged RBL pumps-in charge to the SL. In successive cycles, the voltage on SL increases. As soon as the SL voltage exceeds $V_{REF}$, SA\_P output flips from LOW to HIGH. The number of cycles in the process are counted using a digital counter. On the other hand, for sub-classes SC3 and SC4, SA\_N is enabled with a V$_{REF}$ of $V_{DD}$/2 and 3$V_{DD}$/4, respectively. $PCH_0$ is enabled instead of $\overline{PCH_1}$, thereby pumping-out charge from SL every cycle. Again, the number of cycles are counted when SA\_N flips from HIGH to LOW. This is illustrated in Fig. \ref{fig:steps}, which shows the operation of ADC taking an example of \textit{popcount} cases 0, 8, 24 and 32, for N=64. For the \textit{popcount} case 32 and 24, the sub-classes SC4 and SC3 are determined respectively, thus, charge is pumped out of SL every cycle. Similarly for the \textit{popcount} cases 8 and 0, SC2 and SC1 are determined respectively, and charge is pumped into SL every cycle. The two dummy bitcells are used to mimic the capacitances of the RBLs/RBLBs, such that the charge being pumped in/out from SL every cycle by the dummy bitcells mimics the charge sharing of RBLs/RBLBs and SL in Step 2 (XNOR on SL) operation. Note that the amount of charge being pumped-in/pumped-out exponentially decreases with time. This is a fundamental limit to charge-sharing type ADCs and thus, they work only if the number of counts are small. In our case, for N=64, we count only N/8 = 8 states using this ADC, which gives us fairly accurate results, as shown later. Thus, the output from the first-stage (sub-class SC1-4) along with the output from the second-stage (ADC counts) estimates the number of 1's (\textit{popcount}) for the XNORed output vector. Note that two sets of \textit{popcount}s, one from RWL1a and other from RWL1b, are sequentially read, and then added together to get the final \textit{popcount} of the vector. 

\subsection{Sectioned Memory Array for Parallel Computing}

\begin{figure*}[t]
\centering
\includegraphics[width=\textwidth]{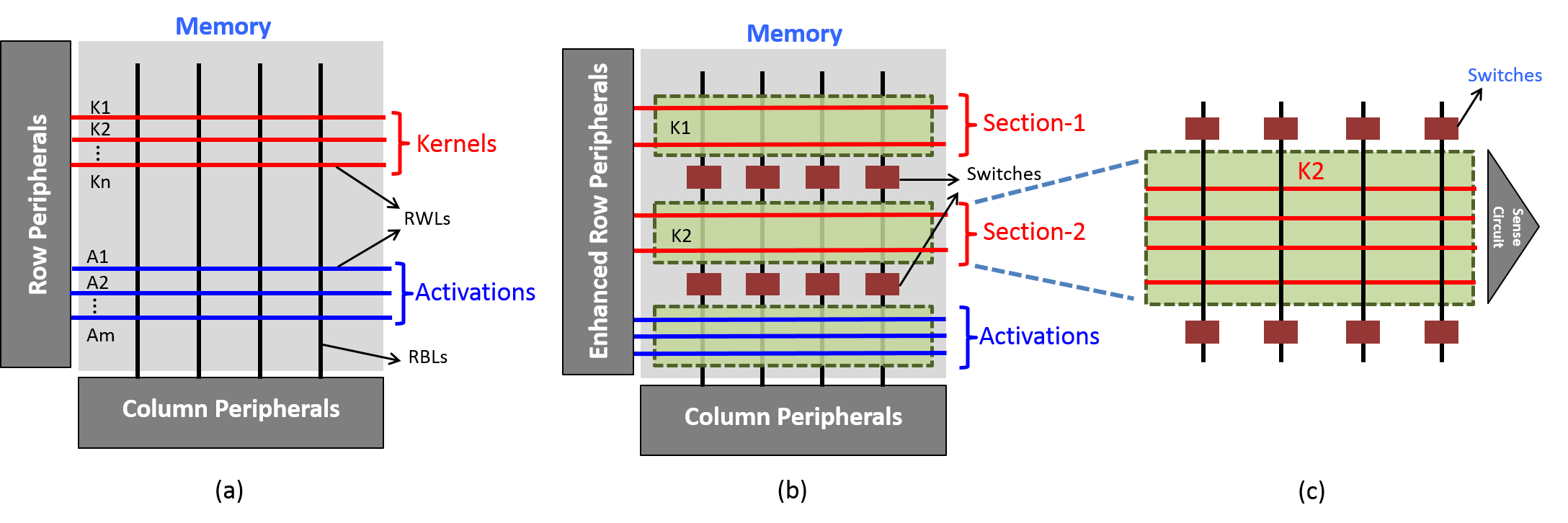}
\caption{a) Typical SRAM memory array with row and column peripherals storing the activations A1-Am, and kernels K1-Kn. b) Proposed sectioned-SRAM array. By introducing switches along the RBLs, the array is divided into sections. The kernels are mapped into the sectioned-SRAM with each section storing different kernel. Once the activations are read onto the RBLs, the switches are opened, and the memory array is divided into sections. c) Since the RBLs for each section have been decoupled, one RWL in each section can be simultaneously enabled such that each section performs the binary convolution concurrently. For example, if A1 was read onto the RBLs before sectioning, enabling the rows K1 and K2 in Section 1 and 2 respectively, we obtain A1*K1 and A1*K2 in parallel.}
\centering
\label{fig:array}
\end{figure*}

We have seen that XNOR and \textit{popcount} operations can be computed within the SRAM array. The manner in which these computations are done, opens possibilities for improving the throughput and energy-efficiency in performing binary convolutions. A typical operation in a CNN layer involves \textit{convolution} of input activations with multiple kernels. This gives us an opportunity for data re-use, since the same set of activations need to be convolved with different kernels. Our proposed scheme described above is well suited to exploit this property of CNNs. Given a set of activations A1, A2,..., Am and kernels K1, K2,..., Kn, stored within the memory array (see Fig. \ref{fig:array}(a)), we need to compute A1*K1, A1*K2,..., A1*Kn, A2*K1, A2*K2,..., A2*Kn and so on. In our computations described above, specifically in the \textit{psedo-read} step, the data corresponding to A1 is read onto the RBL/RBLB voltages. We propose sectioning the memory array into subsections by introducing switches along the RBLs, as shown in Fig. \ref{fig:array}(b), such that kernels are grouped into different sections. Each section consists of a separate ADC control block, as shown in Fig. \ref{fig:array}(c). After A1 has been read onto the RBLs/RBLBs, the switches are opened. The RBLs/RBLBs in individual sections store the information of data A1, but have been decoupled. This allows us to enable one memory row in all the sections corresponding to kernel K1 in section 1, K2 in section 2, and so on, thereby evaluating the XNOR-\textit{popcount} operations concurrently, in all n sections. We thus obtain the output A1*K1, A1*K2,...., A1*Kn in a single cycle. This step can be repeated for all activations A1, A2,..., Am. Thus, sectioning the memory array improves the throughput of our computations n-fold. Moreover, with a single pseudo-read step, we are able to perform \textit{n} convolutions, thereby saving multiple pseudo-read cycles which consume bitline precharge energy. Specifically, without sectioning, one RBL and RBLB pre-charge is required for every convolution operation in addition to ADC energy consumption. With \textit{n}-sections per sub-bank we obtain n-convolutions per pre-charging of the RBL and RBLB thereby not only increasing parallelism but also energy-efficiency.  

Let us now discuss how binary convolutions can be obtained for large kernels using the distributive property of \textit{popcount}. If the kernel size is larger than the memory word length, which is often the case in deeper state-of-the art CNN layers, a single kernel occupies multiple rows in the same or different sub-banks. In-memory binary convolution is performed for each of these kernel rows separately, and the partial \textit{popcount}s obtained from each operation are added to generate the final \textit{popcount}. 
\begin{equation*}
\textit{popcount}(N+N+...)=\textit{popcount}(N)+\textit{popcount}(N)+....
\end{equation*}
Once the final \textit{popcount} is obtained, the output of the binary convolution operation is `1' if the final \textit{popcount} (number of 1's) is greater than half the kernel size, and `0' otherwise.


\subsection{Results}

\begin{figure}[t]
\centering
\includegraphics[width=0.5\textwidth]{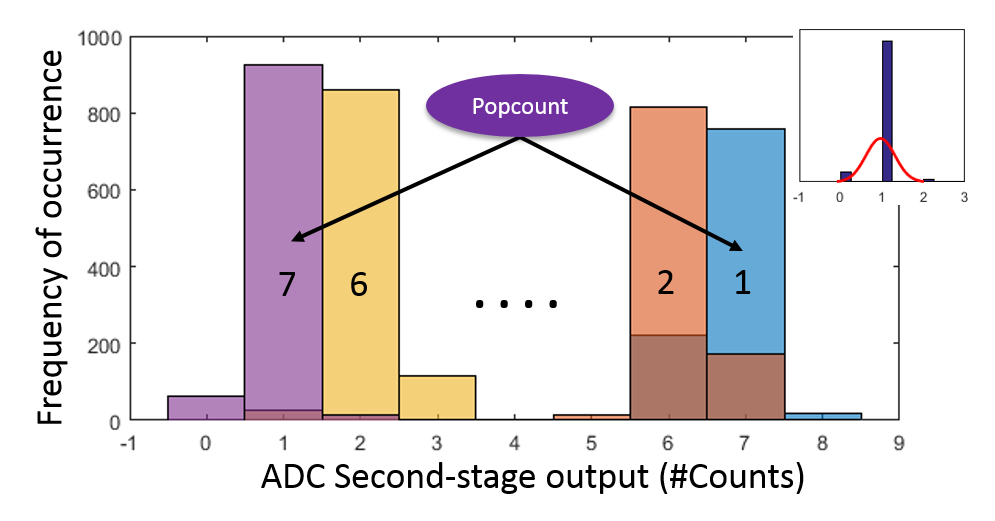}
\caption{Monte-Carlo simulations. The figure plots the histogram of the second-stage output of the ADC, for various \textit{popcount} cases, in presence of process variations. Inset: Each histogram is fitted with a Gaussian distribution. The average standard deviation of the counts is $\sim$0.4359. The trend repeats for higher \textit{popcount} cases with modulo-8, since only the lower 3bits of the output are generated in the second-stage.}
\centering
\label{fig:countdist}
\end{figure}

The sectioned-SRAM array assuming a section size of 32 rows and 64 columns was simulated in HSPICE using the 45-nm predictive transistor models (PTM) \cite{PTM}. As described in the previous section, the final voltage at SL denotes the \textit{popcount} output of the binary convolution. The SL voltage is sensed using the ADC described in the previous section. Again, the 45-nm PTM models were used simulate the SA and the ADC logic block. Using the Dual RWL along with a dual-stage ADC, the ADC output is relaxed to only 5bits. The most-significant bits (2bits) are generated in the first-stage of the ADC (sub-classes SC1-4) using multiple references, while the lower bits (3bits) are generated in the second-stage by the integrating ADC. We observe the effects of CMOS process variation on the ADC output using Monte Carlo simulations, in presence of 30mV sigma threshold voltage variation. Fig. \ref{fig:countdist} plots the distribution of the second-stage ADC output for various \textit{popcount} cases. Note that a similar trend repeats for higher \textit{popcount} cases with modulo-8, since only the lower 3bits of the output are generated in the second-stage. The ADC output is fairly accurate with a small overlap with the neighboring counts. The small inaccuracy is attributed to the transistor threshold voltage variations in the memory array and in the SAs used in the ADC. Moreover, the charge being pumped in/out of SL decreases with each cycle, due to charge-sharing, thereby inducing errors for higher counts. The inset shows a best-fitting normal distribution for the variations in the ADC output. The average standard deviation of the counts was found to be $\sim$0.4359 counts. Total energy consumed per operation was estimated to be $\sim$0.767pJ and $\sim$1.914pJ, with and without sectioned-SRAM (4 sections per bank), respectively. The energy was averaged over various \textit{popcount} cases. Here, by one operation, we mean XNOR + \textit{popcount} of a 64bit input activation and a 64bit kernel, both of which are stored in the SRAM. The energy consumption includes the pre-charge energy in the pseudo-read step and the ADC energy. The latency of one operation was $\sim$45ns. This is due to the low-overhead integrating ADC used, which serially counts to estimate the \textit{popcount}.

\section{In-memory Binary Convolution $-$ Proposal-B}

In the previous section, we described an energy-efficient implementation of performing binary convolutions within the SRAM array. However, the low-overhead ADC used to determine the \textit{popcount} induces errors in the convolution output, which may impact the system accuracy, as we will show later. The primary cause of the inaccuracy is the generation and detection of an analog voltage, which is susceptible to noise, offset etc. Thus, in this section, we propose yet another implementation of enabling binary convolutions in standard SRAM arrays, by modifying the peripheral circuitry. This approach is robust since the \textit{popcount} is computed using digital logic gates (full-adders), unlike Proposal-A which uses analog voltages. Although this robustness comes at a cost of energy-efficiency and throughput as compared to the previous proposal based on charge-sharing, our simulations show that this implementation is still better than the typical von-Neumann based approach as it leverages \textit{in-memory computing} for XNORs and pop-count operations.

\subsection{Bitwise XNORs}

\begin{figure*}[t]
\centering
\includegraphics[width=\textwidth]{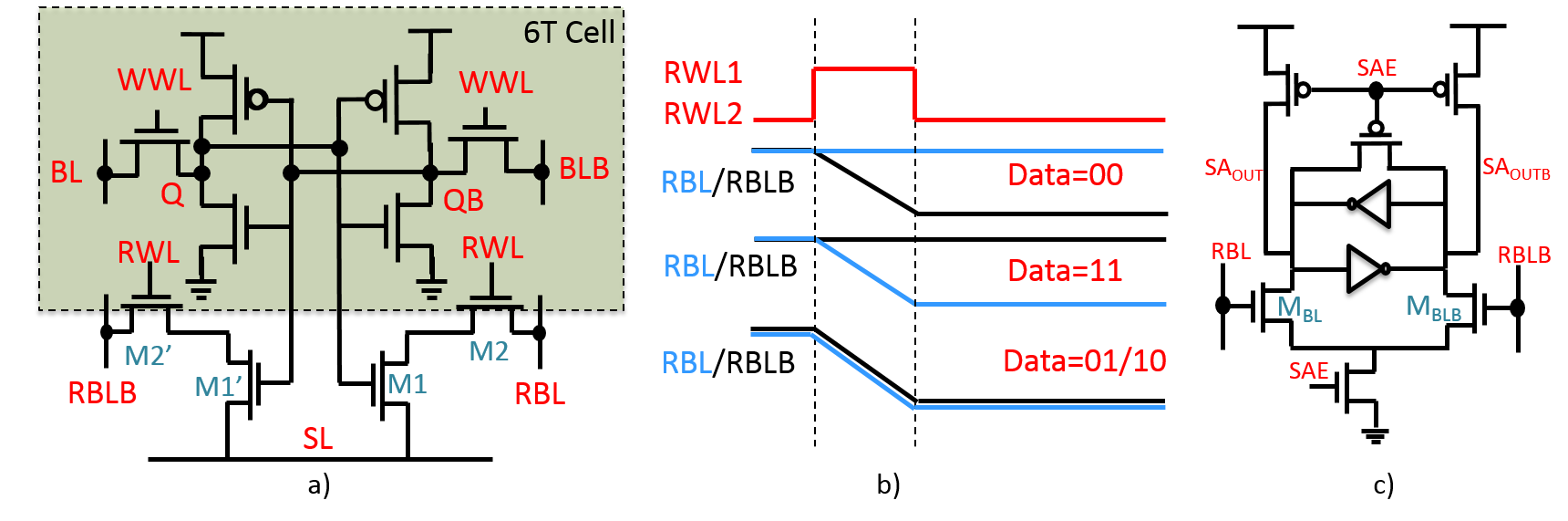}
\caption{(a) A 10T-SRAM bitcell schematic is repeated here for convenience. (b) Timing diagram used for in-memory computing with 10T-SRAM bitcells. (c) Circuit schematic of the asymmetric differential sense amplifier. \cite{agrawal2017x}}
\centering
\label{fig:proposalb}
\end{figure*}

Bitwise Boolean operations within SRAM arrays have recently been demonstrated in \cite{6tddc,rsnm6t,agrawal2017x}. The idea is to enable two RWLs together during a read operation. Let us consider words ‘A’ and ‘B’ stored in two rows of the memory array. Note that we can simultaneously enable the two corresponding RWLs without worrying about read disturbs, since the bit-cell has decoupled read-write paths (shown in Fig. \ref{fig:proposalb}(a)). The RBL/RBLB are pre-charged to $V_{DD}$. For the case ‘AB’ = `00' (`11'), RBL (RBLB) discharges to 0V, but RBLB (RBL) remains in the precharged state. However, for cases `10' and `01', both RBL and RBLB discharge simultaneously. The four cases are summarized in Fig. \ref{fig:proposalb}(b). Now, in order to sense bit-wise XNOR from the RBL/RBLB voltages, we use two asymmetric SAs (see Fig. \ref{fig:proposalb}(c)\cite{agrawal2017x}) which compute the bitwise NAND/NOR in parallel. Asymmetric SAs work by sizing either one of the transistors $M_{BL}$/$M_{BLB}$ bigger than the other. In Fig. \ref{fig:proposalb}(c), if the transistor $M_{BL}$ is sized bigger compared to $M_{BLB}$, its current carrying capability increases. Thus, for cases `01' and `10' where both RBL and RBLB discharge simultaneously, $SA_{out}$ node discharges faster, and the cross-coupled inverter pair of the SA stabilizes with $SA_{out}$=`0'. While for the case `11'(`00'), RBL(RBLB) starts to discharge, and RBLB(RBL) is at V$_{DD}$, making $SA_{out}$=`1'(`0'). Thus it can be observed that $SA_{out}$ generates an AND gate (thus, $SA_{outb}$ outputs NAND gate). Thus, we call this sense-amp $SA_{NAND}$. Similarly, by sizing the $M_{BLB}$ bigger than $M_{BL}$, OR/NOR gates can be obtained and we call it $SA_{NOR}$. Next, by ORing the NOR and AND outputs obtained from $SA_{NOR}$ and $SA_{NAND}$ respectively, bitwise XNOR operation is realized. A detailed description of the bit-wise Boolean XNOR used in this work can be found in \cite{agrawal2017x}.

\subsection{Popcount}

\begin{figure}[t]
\centering
\includegraphics[width=0.5\textwidth]{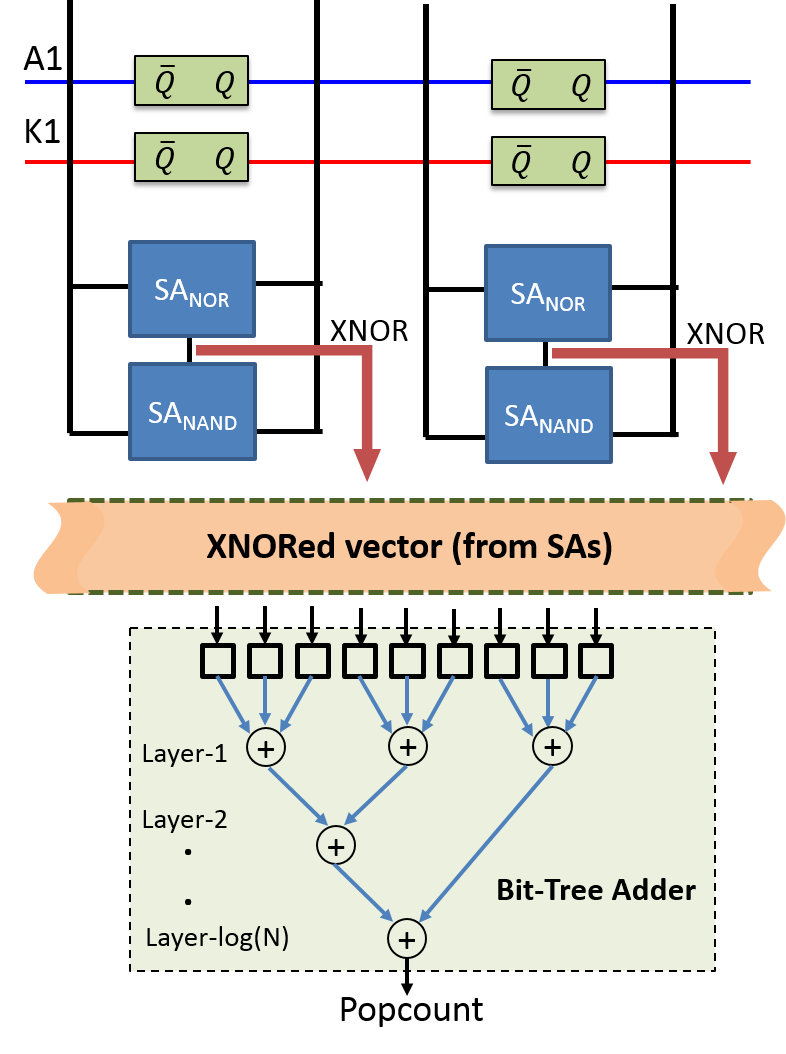}
\caption{Modified peripheral circuitry of the SRAM array to enable binary convolution operations. It consists of two asymmetric SAs - $SA_{NOR}$ and $SA_{NAND}$ which pass the XNORed data vector to a bit-tree adder. The adder has log(N) layers, where N is the number of inputs to the adder. It sums the input bits to generate the \textit{popcount}.}
\centering
\label{fig:bittreeadd}
\end{figure}

In order to utilize the above mentioned approach for enabling binary convolutions, we propose to add a \textit{bit-tree adder} after the asymmetric-SA stage to generate the \textit{popcount}, as shown in Fig. \ref{fig:bittreeadd}. By enabling RWLs corresponding to rows storing activation (A1) and kernel (K1), the asymmetric-SAs generate the XNORed vector. The output XNORed vector is passed to the bit-tree adder. It consists of multiple full-adder (FA) blocks connected in a tree manner. The bit-tree adder sums up all the bits of the output XNORed vector to generate the \textit{popcount}. The first layer of the bit-tree adder consists of single FA blocks, each of which is capable of adding three consecutive bits to generate a 2-bit output. In the next layer, 2-bit adders are used, which are constructed using two stacked FA blocks. The second layer generates 3-bit output. In subsequent layers, multiple FA blocks are stacked to construct multi-bit adders. Finally in the log(N) layer, where N is the number of columns in the sub-array, the \textit{popcount} output is generated, and is read out from the memory. 

To incorporate convolutions with large kernel sizes, the partial \textit{popcount} generated from the bit-tree adders can be summed up over multiple cycles, to generate the final \textit{popcount}. Note that the generated \textit{popcount} is exact, as it is computed using conventional digital logic gates. Also note that the sectioned-SRAM concept described in the previous section is not applicable for this proposal. 

\subsection{Results}


A $128\times64$-bit SRAM array along with the asymmetric SAs $-$ $SA_{NOR}$ and $SA_{NAND}$ were simulated in HSPICE using the 45-nm predictive transistor models (PTM) \cite{PTM}. As described above, two RWLs are enabled simultaneously, and depending on the data stored in each of the bits, $SA_{NOR}$ and $SA_{NAND}$ generate bitwise NOR/OR and NAND/AND, respectively. Readers are referred to \cite{agrawal2017x} for more circuit details and simulations. The energy consumption and latency of the bitwise XNOR operation was estimated to be 29.67fJ/bit and 1ns, respectively. The energy consumption includes the pre-charge energy and the energy consumed in asymmetric-SAs. The bit-tree adder was modeled in Verilog, and synthesized using Synopsys Design Compiler to 45-nm tech node. The inputs to the bit-tree adder block are 64 wires, which represent the bitwise XNORed data generated from the SA stage. The output is a 6-bit \textit{popcount}. The total power and the critical-path delay of the bit-tree adder in performing a 64-bit \textit{popcount}, was estimated to be 0.26mW and 0.3ns, respectively.

\section{System-level Evaluation Framework for BNN}

In this section, we describe the framework developed to evaluate the benefits of our proposals at a system-level, taking an example of a deep binary neural network. We use a modified von-Neumann based system architecture, where the SRAM banks are replaced with our proposed Xcel-RAM banks (Proposal-A/Proposal-B) with embedded convolution compute capabilities. By utilizing these \textit{in-memory} convolutions, we demonstrate the benefits in the overall system energy consumption and latency per inference.

\begin{figure*}[t]
\centering
\includegraphics[width=\textwidth]{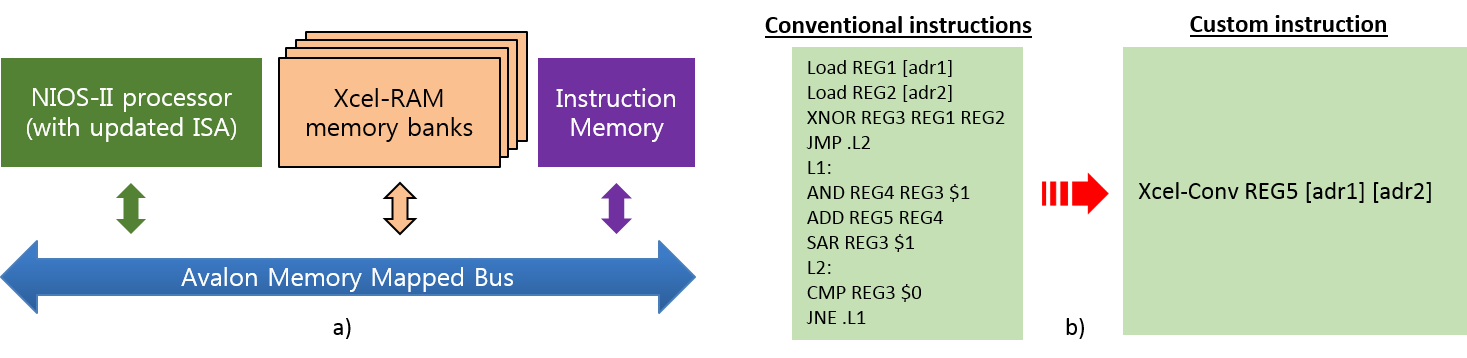}
\caption{(a) Modified von-Neumann architecture based on Xcel-RAM memory banks and enhanced instruction set architecture (ISA) of the processor. (b) Snippet of assembly code for performing a binary convolution operation using conventional instructions and custom instructions.}
\centering
\label{fig:vonneumann}
\end{figure*}

\subsection{Simulation Methodology}


The modified von-Neumann processing architecture is shown in Fig. \ref{fig:vonneumann}(a). It consists of a processor, an Xcel-RAM memory-block and an instruction-memory, connected by a system bus. The Xcel-RAM block consists of multiple subarrays that are arranged in a typical banked structure. We use the CACTI tool \cite{cacti} to model a 64KB Xcel-RAM bank. The circuit numbers for a subarray obtained from HSPICE with the 45nm PTM models \cite{PTM} were put in CACTI to obtain the per-access energy and latency of memory read/write operations as well as binary convolution operation. These include the energy consumed in H-trees, WL decoders, BL drivers, SAs, muxes etc. Next, a cycle-accurate RTL model was developed for Xcel-RAM banks, which was integrated with Intel's programmable Nios-II processor \cite{nios2}, with instruction set (ISA) extensions to leverage the Xcel-RAM compute capabilities (see Fig. \ref{fig:vonneumann}(b)). The system bus follows the Avalon memory-mapped protocol, with enhanced bus architecture to support passing multiple addresses at a time. Note that this is not a large overhead since \textit{in-memory} instructions do not pass the data operands, and thus the data-channel is used to pass extra memory addresses over the bus \cite{sttcim}. Note that although we show a typical von-Neumann based system, Xcel-RAM banks can be interfaced with general purpose graphics processing units (GP-GPUs) based systems as well, to leverage data parallelism along with \textit{in-memory computing}. Our aim here was to show the benefits of replacing conventional SRAM banks with compute capable Xcel-RAM banks.

\begin{table}[t]
\centering
\includegraphics[width=0.5\textwidth]{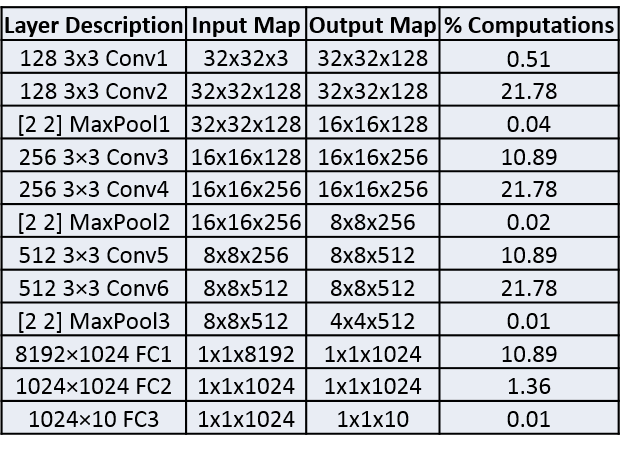}
\caption{Benchmark Binary Neural Network \cite{binarybengio} used for classifying CIFAR10 dataset.}
\centering
\label{fig:bnn}
\end{table}

The binary neural network (BNN) proposed in \cite{binarybengio} uses binary bipolar activations ($\pm1$) for both weights and activations. Note that in our memory, $+1$ is stored as logic HIGH bit, while $-1$ is stored as logic LOW bit. We trained a BNN using the algorithm proposed in \cite{binarybengio} on Pytorch Platform \cite{pytorch} using the github repository \cite{itayhubara2017} of the same work. The neural network architecture is given in Table \ref{fig:bnn}. The network was evaluated on CIFAR-10 \cite{krizhevsky2009learning}. All layers were binarized, except Conv1 and FC3 layers. It was observed that $\sim99.4\%$ of total computations occur in the binarized layers - Conv2-6 and FC1-2, all of which can utilize the Xcel-RAM convolution capabilities (see Table \ref{fig:bnn}). Or in other words, $\sim99.4\%$ of total computations per-inference can be mapped using custom Xcel-RAM instructions, thereby giving us significant improvements in energy and throughput. Each of these layers were run on the modified von-Neumann architecture described above. We assume that the binarized kernels are stored in an off-chip memory, and the kernels for a particular layer are loaded into the SRAM before processing that layer. Typical values of DRAM access energy and latency were taken from literature \cite{Chatterjee_2017}. The software was modified by replacing repetitive convolution operations with our custom instruction macros. In every layer, the convolutions are split into multiple 64-bit XNOR+\textit{popcount} operations, which are then accumulated to compute the final output. The final output is stored back into the SRAM, which would be the input activations for the succeeding layer.

As a baseline, we use a similar system architecture, but with standard SRAM banks with only read/write capability, instead of Xcel-RAM banks. The convolution operation is performed in software through conventional instructions. A snippet of the assembly code for convolution in the baseline and Xcel-RAM based designs is shown in Fig. \ref{fig:vonneumann}(b).

\subsection{Results and Discussion}

\begin{figure*}[t]
\centering
\includegraphics[width=\textwidth]{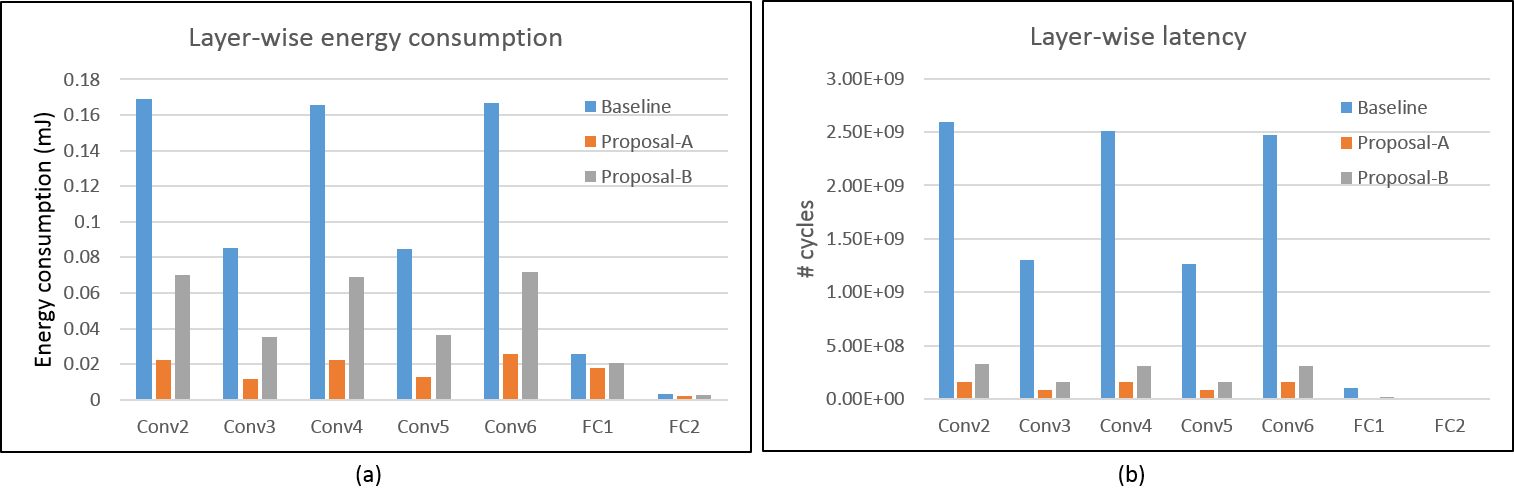}
\caption{(a) Layer-wise (a) energy consumption and (b) latency, for running the CIFAR-10 image classification benchmark on the proposed designs, and the baseline. }
\centering
\label{fig:layerwise}
\end{figure*}

The full precision accuracy of the network was $91.703\%$. The accuracy of the binary neural network was observed to be $89.294\%$, an expected drop due to binarization. We then evaluate the impact of inaccuracies in the ADC for Proposal-A (due to process variations) on the classification accuracy using our simulation framework. At every binarized layer, each element of the output map is a sum of $N$ binary XNORs, where $N = k^2 \times I$, $k$ is the filter height, and $I$ is the number of input channels. Our proposed methodology can perform 64 binary operations at once, in two steps of 32 bits each. Hence, the number of \textit{popcounts} done per element of an output map is $M = ceil(N/32)$. We add the \textit{popcount} error to the output during inference, obtained from circuit simulations, and obtained an accuracy of $88.710\%$, a decrease by $0.584\%$ from the ideal BNN accuracy of $89.294\%$. On the other hand, Proposal-B obtains ideal BNN accuracy because the computations are done using a digital adder-tree.

Fig. \ref{fig:layerwise} shows the layer-wise energy consumption and latency for Proposal-A, Proposal-B, and the baseline. Note that we focus only on layers Conv2-6 and FC1-2, as they constitute majority of the total computations. It can be observed that layers Conv2,4,6 are the most compute intensive layers, due to larger kernels. Overall, per-inference, $6.1\times$ and $2.3\times$ improvements were obtained in energy consumption, for Proposal-A and Proposal-B, respectively, compared to the baseline. 
In terms of latency, $15.8\times$ and $8.1\times$ improvements were obtained per-inference, for Proposal-A and Proposal-B, respectively. These improvements can be attributed to the fact that the most compute intensive operations involved in the BNN inference $-$ bitwise-XNOR followed by \textit{popcount}, are performed efficiently within the memory, thereby saving majority of unnecessary memory accesses and computations. Moreover, the energy and latency benefits of Proposal-A arise from the low-overhead ADC and the sectioned SRAM arrays, which enable multiple operations in a single memory access. In Proposal-B, although the sectioning is not applicable, the energy and latency benefits arise from the bit-wise XNOR computations on the bitline using asymmetric SAs and the digital bit-tree adder to generate the result in the memory array itself.


\section{Conclusion}
Enhanced memory blocks having built-in compute functionality can operate as on-demand accelerators for machine learning computations, while simultaneously operating as usual memory read-write units for general-purpose workloads. In this work, we demonstrated two novel techniques to enable binary convolutions within a standard SRAM memory arrays. In the first proposal, we use charge-sharing on the inherent parasitic capacitances present in the 10T-SRAM structure to embed vector XNOR operations. Further, we use a dual-read wordline along with a dual-stage ADC, to handle the inaccuracies in the low precision, low-overhead ADC. A key highlight of this proposal is the \textit{sectioned-SRAM}, which enables multi-row convolutions in parallel, thereby improving the overall system performance and energy-efficiency. The second proposal uses asymmetric SAs and a bit-tree adder in the memory peripherals to perform bit-wise XNOR computations and \textit{popcount} in-memory. A complete framework was developed to evaluate a benchmark application (CIFAR-10) using our proposed memory arrays. For a system with our proposed Xcel-RAM banks, $6.1\times$ and $2.3\times$ improvements were obtained in energy consumption, and $15.8\times$ and $8.1\times$ improvements were obtained in the latency for the respective proposals, compared to conventional SRAM based system.

\section*{Acknowledgements}
The research was funded in part by C-BRIC, one of six centers in JUMP, a Semiconductor Research Corporation (SRC) program sponsored by DARPA, the National Science Foundation, Intel Corporation and Vannevar Bush Faculty Fellowship.

\bibliographystyle{IEEEtran}
\bibliography{ref}

\end{document}